\documentclass[sigconf]{acmart}

\AtBeginDocument{%
  \providecommand\BibTeX{{%
    \normalfont B\kern-0.5em{\scshape i\kern-0.25em b}\kern-0.8em\TeX}}}

\usepackage{lipsum}
\usepackage{tcolorbox}
\newtcolorbox{mybox}{colback=black!5!white,colframe=black,bottomrule=.25mm,toprule=.25mm,leftrule=.25mm,rightrule=.25mm,left=.25mm,right=.25mm,top=-.25mm,bottom=-.25mm}
   
\usepackage[utf8]{inputenc} 
\usepackage[T1]{fontenc}    
\usepackage{hyperref}      
\usepackage{url}            
\usepackage{booktabs}       
\usepackage{amsfonts}       
\usepackage{nicefrac}       
\usepackage{microtype}      
\usepackage{amsmath}
\usepackage{graphicx}
\usepackage{multirow}
\usepackage{adjustbox}
\usepackage{pifont}
\usepackage{cleveref}
\usepackage{todonotes}
\usepackage{commath}
\usepackage{mathtools}
\usepackage{amsthm}
\usepackage{amssymb}
\usepackage{tikz}
\usepackage{xcolor}
\usepackage{hyperref}

\usepackage[utf8]{inputenc}

\usepackage{color, colortbl}
\definecolor{LightCyan}{rgb}{0.83,0.89,0.75}
\usepackage{graphicx}
\usepackage{adjustbox}
\newtheorem{lemma}{Lemma}
\newtheorem{definition}{Definition}
\newtheorem{theorem}{Theorem}
\usepackage[tight,footnotesize]{subfigure}
\usepackage{doi}
\usepackage[ruled,linesnumbered]{algorithm2e}

\newcommand{\eg}{\textit{e.g\@.}}
\newcommand{\ie}{\textit{i.e\@.}}

\SetKwInput{KwInput}{Input}                
\SetKwInput{KwOutput}{Output}              
\SetKwRepeat{Do}{do}{while} 
\usepackage{mathtools, nccmath}
\usepackage{wrapfig}
\usepackage{comment}
\usepackage{graphicx}
\usepackage{xspace}
\usepackage{array}
\usepackage{ragged2e}
\newcolumntype{P}[1]{>{\RaggedRight\hspace{0pt}}p{#1}}
\newcolumntype{X}[1]{>{\RaggedRight\hspace*{0pt}}p{#1}}
\usepackage{tcolorbox}
\usepackage{tikz}
\usetikzlibrary{arrows,shapes,positioning,shadows,trees,mindmap}
\usepackage[edges]{forest}
\usetikzlibrary{arrows.meta}
\colorlet{linecol}{black!75}
\usepackage{xkcdcolors} 
\usepackage{tikz}
\usetikzlibrary{backgrounds}
\usetikzlibrary{arrows,shapes}
\usetikzlibrary{tikzmark}
\usetikzlibrary{calc}

\newenvironment{sloppypar*}
{\sloppy\ignorespaces}
{\par}
\AtBeginDocument{%
  \providecommand\BibTeX{{%
    \normalfont B\kern-0.5em{\scshape i\kern-0.25em b}\kern-0.8em\TeX}}}

\usepackage[framemethod=TikZ]{mdframed}
\usepackage[flushleft]{threeparttable}
\usepackage{caption}
\usepackage{tikz}

\copyrightyear{2024}
\acmYear{2024}
\setcopyright{acmlicensed}
\acmConference[KDD '24] {Proceedings of the 30th ACM SIGKDD Conference on Knowledge Discovery and Data Mining }{August 25--29, 2024}{Barcelona, Spain.}
\acmBooktitle{Proceedings of the 30th ACM SIGKDD Conference on Knowledge Discovery and Data Mining (KDD '24), August 25--29, 2024, Barcelona, Spain}
\acmISBN{979-8-4007-0490-1/24/08}
\acmDOI{10.1145/3637528.3671879}

\usepackage[utf8]{inputenc}
\usepackage{colortbl}

\definecolor{auburn}{rgb}{0.43, 0.21, 0.1}
\definecolor{burgundy}{rgb}{0.5, 0.0, 0.13}

\newcommand{\tabincell}[2]{\begin{tabular}{@{}#1@{}}#2\end{tabular}}
\newtheorem{assumption}{Assumption}

\begin{document}\sloppy

\title{\textit{\textsf{BadSampler}}: Harnessing the Power of Catastrophic Forgetting to Poison Byzantine-robust Federated Learning}

\author{Yi Liu}
\affiliation{%
  \institution{City University of Hong Kong}
  \city{Hong Kong}
  \country{China}
}
\email{yiliu247-c@my.cityu.edu.hk}

\author{Cong Wang}
\authornote{Corresponding author. This work was supported by CityU of HK under Grants 9678146 and 9678126, in part by HK RGC under Grants CityU 11218521, 11218322, R6021-20F, R1012-21, RFS2122-1S04, C2004-21G, C1029-22G, and N\_CityU139/21.}
\affiliation{%
  \institution{City University of Hong Kong}
  \city{Hong Kong}
  \country{China}
}
\email{congwang@cityu.edu.hk}

\author{Xingliang Yuan}
\affiliation{%
  \institution{The University of Melbourne}
  \city{Melbourne}
  \country{Australia}
}
\email{xingliang.yuan@unimelb.edu.au}

\renewcommand{\shortauthors}{Yi Liu et al.}

\begin{abstract}
Federated Learning (FL) is susceptible to poisoning attacks, wherein compromised clients manipulate the global model by modifying local datasets or sending manipulated model updates. Experienced defenders can readily detect and mitigate the poisoning effects of malicious behaviors using Byzantine-robust aggregation rules. However, the exploration of poisoning attacks in scenarios where such behaviors are absent remains largely unexplored for Byzantine-robust FL. This paper addresses the challenging problem of poisoning Byzantine-robust FL by introducing catastrophic forgetting. To fill this gap, we first formally define generalization error and establish its connection to catastrophic forgetting, paving the way for the development of a clean-label data poisoning attack named \emph{\textsf{BadSampler}}. This attack leverages only clean-label data (\ie, without poisoned data) to poison Byzantine-robust FL and requires the adversary to selectively sample training data with high loss to feed model training and maximize the model's generalization error. We formulate the attack as an optimization problem and present two elegant adversarial sampling strategies, Top-$\kappa$ sampling, and meta-sampling, to approximately solve it. Additionally, our formal error upper bound and time complexity analysis demonstrate that our design can preserve attack utility with high efficiency. Extensive evaluations on two real-world datasets illustrate the effectiveness and performance of our proposed attacks.
\end{abstract}

\ccsdesc[500]{Security and privacy~Distributed systems security}
\ccsdesc[500]{Computing methodologies~Machine learning}

\keywords{Federated Learning, Data Poisoning Attack, Clean Lable, Reinforcement Learning}

\maketitle
\section{Introduction}
Federated Learning (FL)~\cite{mcmahan2017communication}, as an emerging distributed machine learning framework, has provided users with many privacy-friendly promising applications~\cite{meng2021cross,zheng2022aggregation,zhang2020enabling,rahman2023fedpseudo,wu2023serverless}. However, it is known that FL is vulnerable to \emph{poisoning}~\cite{tolpegin2020data,shejwalkar2021back,shejwalkar2021manipulating}: a small fraction of \emph{compromised} FL clients, who are either owned or controlled by an adversary, may act maliciously during the training process to corrupt the jointly trained global model, \ie, degrading global model performance or causing misclassification in prediction. In this context, various attack strategies aimed at perturbing local training datasets or generating poisoned gradients have been proposed, so-called \emph{Data Poisoning Attacks (DPA)}~\cite{shafahi2018poison} or \emph{Model Poisoning Attacks (MPA)}~\cite{shejwalkar2021back}. However, prior arts~\cite{shejwalkar2021back,wangmore} have shown that existing attacks are difficult to deploy in practice because they either rely on unrealistic assumptions (\eg, manipulating too many clients), or the effectiveness of these attacks is greatly suppressed by state-of-the-art defensive Byzantine-robust aggregation rules (\eg, \textsf{FLTrust}~\cite{caofltrust}).

In this paper, we endeavor to investigate the FL poisoning problem in a more practical scenario, where the adversary needs to achieve successful poisoning of FL without resorting to establishing unrealistic attack assumptions or deploying advanced defensive aggregation rules. In addition, aligning with existing work~\cite{caofltrust,jagielski2018manipulating,lin2019byzantine}, we likewise need to push the frontiers of FL poisoning attacks by considering the deployment of strong defenses. In particular, we target Byzantine-robust FL~\cite{caofltrust,mao2021romoa}, which represents a widely embraced collaborative learning framework known for its resilience against attacks. In the Byzantine-robust FL paradigm, the central server employs robust aggregation rules for the purpose of model update aggregation. This facilitates the filtering of poisoned updates or malicious gradients, thereby preserving the overall model utility~\cite{caofltrust,awan2021contra,lin2019byzantine}. As a result, implementing poisoning attacks in the context of Byzantine-robust FL without relying on unrealistic attack assumptions presents a formidable challenge.

\noindent \textbf{Our Contributions.} To tackle the aforementioned challenges, we present \emph{\textsf{BadSampler}}, which stands as an effective clean-label data poisoning attack designed exclusively for Byzantine-robust FL. Unlike existing attacks, \emph{\textsf{BadSampler}} leverages only clean-label data to achieve its goals. In the FL training pipeline, the widely used Stochastic Gradient Descent (SGD) typically employs uniform random sampling of training instances from the dataset. However, this randomness is seldom tested or enforced in practical implementations~\cite{shejwalkar2021manipulating,smith2020origin}. This situation presents an opportunity for adversaries to exploit the vulnerability, allowing them to devise an adversarial sampler that induces catastrophic forgetting (a phenomenon where models lose information from previous tasks while learning new ones~\cite{kirkpatrick2017overcoming}) in the FL model. \emph{\textsf{BadSampler}} is specifically designed to address this issue, with two primary goals in mind: \emph{(1)} adaptively select training samples to poison FL, causing catastrophic forgetting while maintaining good training behavior, and \emph{(2)} bypass the latest defenses within a practical FL parameter range. Next, we focus on responding to the following three challenges:

$\bullet$ \textit{(C1.) How to introduce catastrophic forgetting under Byzantine robust aggregation rules?}

\noindent \textbf{\textit{(S1.)}} -- \textit{Not all benign updates are beneficial.} In FL, benign training behavior means that the updates generated by iteration are beneficial to the model to obtain high training accuracy~\cite{mcmahan2017communication}. Inspired by the phenomenon of catastrophic forgetting~\cite{kirkpatrick2017overcoming}, well-designed benign updates (\ie, achieving higher training accuracy) may cause the global model to forget knowledge about old tasks, thus damaging the model's generalization ability. Furthermore, prevailing Byzantine robust aggregation protocols are limited to mitigating the adverse effects stemming from adversarial training behaviors, specifically poisoned updates and gradients~\cite{caofltrust,awan2021contra}, while encountering challenges in discerning the nature of benign training behaviors. Therefore, \emph{\textsf{BadSampler}} is designed to circumvent advanced defenses by maintaining good training behavior. The key to maintaining good training behavior is to maintain a low training error but slowly increase the validation error. To achieve this, the proposed attack strategy endeavors to accomplish high-frequency sampling of training samples characterized by elevated losses through the construction of adversarial samplers. This approach is designed to uphold the model's attainment of high training accuracy while concurrently inducing catastrophic forgetting phenomena.

$\bullet$ \textit{(C2.) How to design an effective poisoning attack under strict parameter constraints?}

\noindent \textbf{\textit{(S2.)}} -- \textit{Not all poisoning attacks are subject to parameter constraints.} It is also known that most of the prior poisoning attacks are less effective under a practical FL parameter range~\cite{shejwalkar2021back,wangmore}, such as a realistic compromised client ratio ($M<20\%$), participation ratio ($q=10\%$), and data poisoning size ($|D_{poison}| \gg |D|$). Under these settings, enabling a stealthy poisoning attack becomes even more difficult. For this reason, \emph{\textsf{BadSampler}} is designed to no longer use poisoned data but to use clean-labeled data to implement the attack. More specifically, \emph{\textsf{BadSampler}} can cause catastrophic forgetting in the global model simply by changing the order in which local clean data is fed to the model. Furthermore, \emph{\textsf{BadSampler}} does not rely on changing parameter settings to improve attack performance. The key information that our attack relies on is the generalization error representation (refer to Sec. \ref{sec-3-2}), so it is data-agnostic and model-agnostic and independent of external FL parameters (\ie, controlled by the server). Furthermore, our theoretical analysis confirms that the error upper bound of our attack is related to the internal FL parameters (\ie, controlled by the clients).

$\bullet$ \textit{(C3.) How to maintain excellent attack effectiveness under FL's dynamic training principles?}

\noindent \textbf{\textit{(S3.)}} -- \textit{Heuristic poisoning attacks are feasible.} FL's dynamic training principle is reflected in the way that the server re-selects different clients to participate in training each iteration~\cite{mcmahan2017communication}. Existing poisoning attacks often require compromising a fixed set of clients to participate in each round of training rather than randomly selecting available clients as per the FL training pipeline. Such a requirement heavily reduces attack practicality. \emph{\textsf{BadSampler}} demonstrates adaptability by enabling different compromised clients at each training iteration to leverage the training state, \eg, error information, to make effective adversarial sampling decisions. The attack formulates the attack target as an optimization problem and leverages the Top-$\kappa$ and meta-sampling strategies to approximately solve it. Moreover, our attack formalizes a model-agnostic generalization error representation, and the objective function differs from the main task, allowing optimization in each iteration.  

The contributions of this paper are listed as follows:

\emph{(1)} We design a new clean-label data poisoning attack, \ie, \emph{\textsf{BadSampler}}, against Byzantine-robust FL models under practical parameter range constraints by introducing catastrophic forgetting. 

\emph{(2)} We design two optimizable adaptive adversarial sampling strategies to maintain the utility of the proposed attack in practical FL scenarios.

\emph{(3)} We conduct extensive experiments on two public datasets for convex, and non-convex models, and evaluate over prior defenses to demonstrate the advantages of the proposed attacks. In particular, our attack can lead to a $8.98\%$ drop in the accuracy of \textsf{FLTrust}~\cite{caofltrust}) compared to the baselines.

\section{Related Work}\label{sec-2}
\noindent \textbf{Existing Poisoning Attacks and Their Limitations.} 
The current poisoning attacks in FL primarily follow two common strategies: \emph{(1)} Directly modifying the local data~\cite{tolpegin2020data,jagielski2018manipulating,jere2020taxonomy}; \emph{(2)} Building adversarially-crafted gradients~\cite{shejwalkar2021manipulating,ref-13,ref-15}. Although these attack tactics indeed pose real threats, they have evident limitations in practical scenarios. Here we explore the limitations of these attack strategies in a more realistic FL scenario where training operates within a practical range of parameters (refer to Table \ref{tab:practical_ranges}). Firstly, DPAs necessitate the adversary possessing knowledge of the dataset or model and the capability to perturb the training dataset~\cite{shejwalkar2021back,shejwalkar2021manipulating}. Fulfilling this requirement proves challenging in practice. Experienced defenders, especially those equipped with anomaly detection techniques, can swiftly identify corrupted datasets~\cite{ref-75,awan2021contra}. Moreover, studies have shown that DPAs struggle to bypass well-designed defenses, such as Byzantine robust aggregation rules~\cite{caofltrust,mao2021romoa}, even if the adversary successfully acquires dataset and model knowledge while circumventing the anomaly detection mechanism. Secondly, existing MPAs heavily rely on sophisticated and high-cost joint optimization techniques among compromised clients, as described in references~\cite{ref-13,ref-15}. These attacks also rely on irrational parameter values, such as an excessive percentage of compromised clients (typically $> 20\%$)~\cite{ref-13}. These limitations motivate us to explore a poisoning attack that does not require dataset perturbation, utilizes a clean-label dataset, and bypasses Byzantine robust aggregation.

\noindent \textbf{Defenses Against Poisoning Attacks in FL.} Existing defenses against poisoning attacks can be categorized into two main strategies. Firstly, there are defenses aimed at detecting and removing malicious clients with anomalous data or model updates~\cite{ref-72,ref-78,sun2023shapleyfl,yan2023criticalfl,zhang2022fldetector}. Secondly, there are defenses aimed at limiting the negative impact of such attacks, such as using Byzantine-robust aggregation methods that aggregate local model parameters using the Geometric mean instead of the mean~\cite{caofltrust,park2021sageflow,lin2019byzantine,yin2018byzantine}. In this paper, our focus is on Byzantine robust FL, and these defenses serve as commonly used methods against both data poisoning attacks and model poisoning attacks~\cite{yin2019defending}. Specifically, the goal of poisoning attacks is to manipulate a limited number of malicious clients to poison the aggregated model, resulting in an inevitable drop in model accuracy. To counter this, the key idea behind Byzantine-robust Aggregation (AGR) is to identify and remove anomalous updates before updating the aggregated model by comparing the clients' local model updates. For instance, Li \textit{et al.} in \cite{ref-76} proposed a new aggregation rule termed \textsf{RSA}, which employs the geometric median of the client's local update as the global model update. The current state-of-the-art defense in this context is \textsf{FLTrust}~\cite{caofltrust}, which leverages trust scores to determine which model updates can be aggregated.

\section{Background and Threat Model}
\subsection{Definition of the Generalization Error}\label{sec-3-2}
Here, we focus on elaborating on the formal definition of generalization error. In the machine learning domain, the bias-variance trade-off is the basic theory for qualitative analysis of generalization error~\cite{geman1992neural,liu2021impact,zhou2023understanding}, which can be defined as follows:
\begin{definition}\label{defi-1} (Generalization Error). Let $D =\{ \boldsymbol{x},\boldsymbol{y}\} =\{ {x_i},{y_i}\} _i^n$ that contains $n$ training samples denote the training dataset, where $x_i$ denotes the $i$-th training sample, and $y_i$ is the associated label. Let $f_\theta(x)$ denote the predictor and let $\ell ( \cdot , \cdot )$ denote the loss function. Thus, the formal expression of the generalization error is as follows:
\begin{equation}\label{eq-1}
    \begin{aligned}
\text { Err } & =\mathbb{E}_{\boldsymbol{x}, \boldsymbol{y}} \mathbb{E}_D\left[\|\boldsymbol{y}-f_\theta(\boldsymbol{x}|D)\|_2^2\right] \\
& \approx \underbrace{\mathbb{E}_{\boldsymbol{x}, \boldsymbol{y}}\left[\|\boldsymbol{y}-\bar{f}_\theta(\boldsymbol{x})\|_2^2\right]}_{\text {Bias}}+\underbrace{\mathbb{E}_{\boldsymbol{x}, \boldsymbol{y}} \mathbb{E}_D\left[\|f_\theta(\boldsymbol{x}|D)-\bar{f}_\theta(\boldsymbol{x})\|_2^2\right]}_{\text {Variance}},
\end{aligned}
\end{equation}
where $\bar{f}_\theta(\boldsymbol{x})=\mathbb{E}_D[f_\theta(\boldsymbol{x}|D)]$. According to the above definition, the bias may be smaller when the variance is larger, which means that the training error of the model is small and the verification error may be large. When the above situation occurs, the model is likely to fall into the catastrophic forgetting problem.
\end{definition}
\begin{definition}\label{defi-2}
    (Generalization Error of Sample $x_i$).  Similarly, we define the generalization error of a single sample $x_i$ as
    \begin{equation}\label{eq-2}
    \operatorname{err}_i=\mathbb{E}_D\left[\ell\left(f_\theta\left({x}_i | D\right), y_i\right)\right] \approx Bias\left({x}_i\right)+Variance\left({x}_i\right),
\end{equation}
where $Bias\left({x}_i\right)$ and $Variance\left({x}_i\right)$ are the bias and variance of $x_i$. Obviously, when the generalization error of the sample $x_i$ approaches $1$, we can call it a hard sample; on the contrary, we can call the sample $x_i$ an easy sample. Therefore, for the sample $x_i$, \textit{the definition of the generalization error can be used to represent the difficulty of sample learning}, \ie, the sample difficulty~\cite{geman1992neural,liu2021impact,zhou2023understanding}.
\end{definition}

\subsection{Threat Model}
We consider a threat model in which an adversary could exploit well-crafted adversarial samplers to poison the global model. During training, the adversary forges a sampler that embeds a malicious sampling function. This adversarial sampler is then incorporated into the target FL system through system development or maintenance~\cite{shumailov2021manipulating}. At inference time, the target FL system has indiscriminately high error rates for testing examples. Next, we discuss our threat model with respect to the attacker’s goals, background knowledge, and capabilities.

\noindent \textbf{Attacker’s Goal.} The attacker's goal is to manipulate the learned global model to have an indiscriminately high error rate for testing examples, which is similar to prior studies on poisoning attacks~\cite{ref-78,ref-13,ref-75,ref-63}. Given that there are many sophisticated defenders against poisoning attacks today~\cite{caofltrust,rieger2022deepsight}, the attacker's second goal is to successfully evade the tracks of these defenders. 

\noindent \textbf{Attacker’s Background Knowledge.} We assume that the attacker holds a model (called \emph{surrogate model}) that is not related to the primary task. It should be noted that the surrogate model is utilized by the attacker to build an adversarial sampler but has no impact on the training of the primary task. Furthermore, to make the attack more stealthy and practical, we assume that the attacker only needs to access and read the local dataset. Although previous literature~\cite{shejwalkar2021manipulating,shafahi2018poison,bonawitz2019towards} has shown that attackers can modify and add perturbations to local datasets, our experimental results show that such operations are easily detected by experienced defenders.

\noindent \textbf{Attacker’s Capability.} We assume that the attacker controls the proportion $M$ of all clients $K$, called compromised clients. Following the production FL settings (see Appendix \ref{sec-3-1}), we assume that the number of compromised clients is much less than the number of benign clients, \ie, $M \leqslant 10\%$. 

\noindent \textbf{Remark.} We provide the following justifications for why our attack is realistic and distinct from other poisoning attacks: \emph{(1)} In the context of FL, the client possesses autonomous control over the local training process, rendering it feasible for a compromised client to deceitfully forge an adversarial sampler, thereby manipulating the local training dynamics. \emph{(2)} The client is solely required to have read permission for the local dataset and does not need write permission. As read operations generally necessitate less privilege than write operations, this approach is practical and helps preserve the cleanliness of the dataset by avoiding any modifications to it.

\section{BadSampler Attack}
\subsection{Primer on BadSampler Attack}\label{sec-3-2}
Following \cite{shejwalkar2021manipulating}, we first demonstrate that the order of local training batches provided to model training affects model behavior. In FL, the central server and clients cooperatively run a $T$-round training protocol to perform federated optimization. Referring to FedAvg~\cite{mcmahan2017communication}, the formal definition of federated optimization is as follows:
\begin{equation}\label{eq-3}
\vspace{-0.3cm}
\mathop {\min }\limits_\omega f(\omega ),{\text{where}}\text{ }f(\omega ): = \sum\limits_{k = 1}^K {{p_k}{F}_k(\omega)},
\end{equation}
where ${p_k} \geqslant 0,\sum\nolimits_k {{p_k} = 1} $ is a user-defined term that indicates the relative influence of each client on the global model. In Eq. \eqref{eq-3}, we mainly focus on the \emph{local optimization} term $F_k(\omega)$ which determines the performance of the global model. In fact, this local optimization term can be regarded as solving a \emph{non-convex optimization problem} with respect to the parameter $\omega^k$, corresponding to the minimization of a given local loss function $F_k(\omega^k)$. For the $k$-th client, we consider that the batch size of local training is $B$ and let $N \cdot  B$ be the total number of items for training, then in a single epoch one aims to optimize: ${F_k}({\omega ^k}) = \frac{1}{N}\sum\nolimits_{i = 1}^N {{\ell _i}(} f_k({x_i}),y_i)$. 

In general, we use the SGD algorithm to solve the above optimization problem, \ie, the following weight update rule is implemented in a training epoch: 
$\omega _{t + 1}^k = \omega _t^k + \eta \Delta \omega _t^k,\Delta \omega _t^k =  - {\nabla _\omega }\ell ({\omega ^k}),$ where $\eta$ is the learning rate.
Secondly, we delve into the sampling procedure of the SGD algorithm and analyze the impact of its batching on model performance. In fact, the stochasticity of SGD comes from its sampling process, thus, in SGD, sampling affects how well the mini-batch gradient approximates the true gradient. Given the local data $D_k$, we assume an unbiased sampling procedure, and the expectation of the batch gradient matches the true gradient as follows:
\begin{equation}\label{eq-4}
\mathbb{E}[\nabla \ell _{{i_t}}^k({\omega ^k})] = \sum\limits_{i = 1}^N {\mathbb{P}({i_t} = i)\nabla } \ell _i^k({\omega ^k}) = \nabla {F_k}({\omega ^k}),
\end{equation}
where ${\mathbb{P}({i_t} = i)}=\frac{1}{N}$ and $t$ is the number of SGD steps.

\begin{figure*}[!t]
 \centering
 \includegraphics[width=1\linewidth]{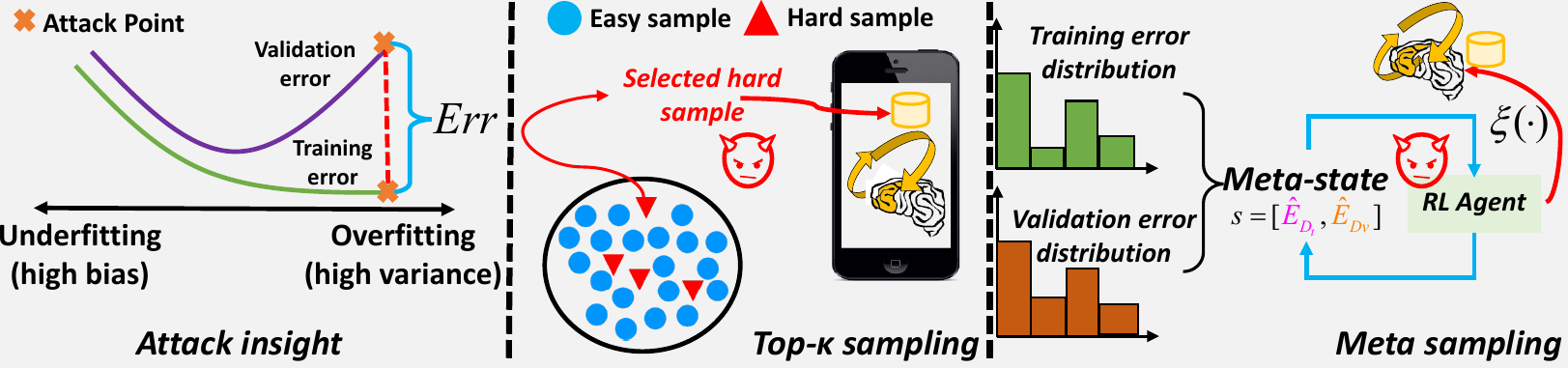}
  \caption{Workflow and taxonomy of our \emph{\textsf{BadSampler}} attack.}
  \label{fig-1}
\end{figure*}

\noindent \textbf{Observations.} We emphasize that Eq. \eqref{eq-4} occurs only as expected, and conversely, for isolated training batches, the approximation error may be large and affect model performance. Inspired by this fact, we investigate how adversaries exploit local sampler to disrupt training. This means that the classical stochastic assumption in SGD opens up a new attack surface for adversaries to affect the learning performance and process~\cite{shumailov2021manipulating,smith2020origin}. To this end, we revisit the impact of $N$ SGD steps during local training on model performance:
\begin{equation}
            \label{eq-5}
\begin{aligned}
  \omega _t^k &= \omega _0^k - \eta \nabla {\ell _0}(\omega _0^k) - \eta \nabla {\ell _1}(\omega _1^k) - \eta \nabla {\ell _2}(\omega _2^k) -  \cdots  \\
   &= \omega _0^k - \eta \sum\limits_{j = 0}^{N - 1} \nabla  {\ell _j}(\omega _0^k) + {\eta ^2}\sum\limits_{j = 0}^{N - 1} {\sum\limits_{i < j} {\nabla \nabla } } {\ell _j}(\omega _0^k){\ell _i}(\omega _0^k)+ \mathcal{O}({N^3}{\eta ^3}) \\
   &= \omega _0^k - N\eta \nabla \ell (\omega _0^k) + \eta ^2\zeta (\omega _0^k) +\mathcal{O}({N^3}{\eta ^3}),
\end{aligned}
\end{equation}
where $\mathcal{O}({N^3}{\eta ^3})$ is the error caused by the Taylor expansion and second order correction term $\zeta (\omega _0^k) = \sum\limits_{j = 0}^{N - 1} {\sum\limits_{i < j} {\nabla \nabla } } {\ell _j}(\omega _0^k){\ell _i}(\omega _0^k)$ is the stochastic error determined by the mini-batches. This means that we can simply manipulate local training mini-batches to harm learning performance and convergence. Hence, our poisoning attack aims to manipulate local samplers to change the order of the local training batch without using poisoned data. By doing so, we cause the first and second derivatives to be misaligned with the true gradient step, thereby undermining the model's generalization ability.

\subsection{Formulating Optimization Problems}
Our idea is to poison Byzantine-robust FL by utilizing adversarial training batches sampled by a carefully crafted adversarial sampler at each iteration. More specifically, we aim to use local training batches sampled by adversarial samplers to increase the generalization error of the FL global model so that FL falls into catastrophic forgetting. Therefore, we expect to find a model-agnostic representation that can provide generalization error information for adversarial samplers. Inspired by the theory of bias-variance tradeoff (see Defi. \ref{defi-1}), the attacker's goal translates to performing adversarial sampling among compromised clients to obtain the most appropriate adversarial training batches to maximize Eq. \eqref{eq-1}. Without loss of generality, we consider that the first $m$ clients are compromised. Our generalization error maximization objective is to craft locally adversarial training batches $\mathcal{B}_{1}^{\prime},\mathcal{B}_{2}^{\prime}, \cdots ,\mathcal{B}_{m}^{\prime}$ for the compromised clients via solving the following optimization problem at each iteration:
\begin{align}\label{eq-7}
&\max_{\mathcal{B}_{1}^{\prime},\mathcal{B}_{2}^{\prime}, \cdots ,\mathcal{B}_{m}^{\prime}}  \text { Err },\nonumber \\
\text{subject to } & \mathcal{B}^{\prime}= \mathcal{A}(\mathcal{B}_{1}^{\prime},\mathcal{B}_{2}^{\prime}, \cdots ,\mathcal{B}_{m}^{\prime}), \nonumber \\
&(x,y) \in D_m, \nonumber \\
&\mathop {\min }\limits_{(x,y) \in {D_m}} \mathbb{E}_{{D_t}}^m,
\end{align}
where $\mathcal{B}$ denotes that the after-attack adversarial training batch on the compromised client, $(x,y) \in D_m$ denotes that only clean-label data can be used, and $\mathop {\min }\limits_{(x,y) \in {D_m}} \mathbb{E}_{{D_t}}^m$ denotes that the training error is minimized (aim to bypass advanced defenses). From the above optimization goals, it can be seen that the attacker needs to maximize the generalization error on the one hand and minimize the training error on the other to prevent being filtered by experienced defenders.

\subsection{Attack Implementation}
The attacker's goal is formulated as an optimization problem in Eq. \eqref{eq-7}. However, due to the non-convex and nonlinear nature of Eq. \eqref{eq-7}, solving it exactly is impractical. To make it amenable to optimization, we propose an effective adversarial sampler called \emph{\textsf{BadSampler}}, which guides hard sampling to optimize Eq. \eqref{eq-7}. The workflow and taxonomy of \emph{\textsf{BadSampler}} attacks are illustrated in Fig. \ref{fig-1}. The proposed attack strategies consist of \emph{(1)} the top-$\kappa$ sampling strategy and \emph{(2)} the meta-sampling strategy. The core idea of the first attack strategy is to construct a training sample candidate pool with high loss and sample the hard training samples from this pool to feed the model. The insight of the second attack strategy is to treat the training and validation error distribution as the meta-state of the error optimization process, thus using meta-training to maximize the generalization error.

\noindent \textbf{Top-$\kappa$ Sampling Strategy.} As mentioned above, the core idea of this strategy is the construction of a hard sample candidate pool. To select those samples that are conducive to maximizing the generalization error, we use the sample difficulty (see Defi. \ref{defi-2}) as a selection metric to select appropriate samples. Specifically, we select the top $\kappa*B$ ($\kappa$ is a constant and $B$ is the size of the training batch) difficult-to-learn samples as the hard sample candidate pool and instruct the adversarial sampler to sample data from this candidate pool to feed the model training. \textit{Note that sample difficulty is calculated by the surrogate model because sample difficulty is model-agnostic and independent of dataset size and sample space~\cite{shejwalkar2021manipulating}.} For simplicity, let $\mathcal{H}_{bad}$ denote the hard sample pool, then the selection process can be formally defined as follows:
\begin{equation}\label{eq-8}
{\mathcal{H}_{bad}} = Top - \kappa (er{r_i},{D_m},B).
\end{equation}
Then the process of sampling hard samples from $\mathcal{H}_{bad}$ to feed the model training can be expressed as: $f({\theta},{x_i} |{\mathcal{H}_{bad}})$. Note that we resample the samples in this candidate pool and the candidate pool is also updated iteratively.

\noindent \textbf{Meta Sampling Strategy.} In this strategy, we adopt the concept of \emph{meta-learning} to seek a model-agnostic representation that provides generalization error information for adversarial meta-samplers. By leveraging the bias-variance trade-off, we can express the generalization error in a form independent of task-specific properties, such as dataset size and feature space~\cite{yang2020rethinking,amin2019bounding}. Consequently, this representation enables adversarial samplers to perform across different tasks. Drawing inspiration from Definition 1 and the idea of ``gradient/hardness distribution''~\cite{geman1992neural,liu2020mesa,liu2020self}, we introduce the histogram distribution of training and validation errors as the \emph{meta-state} of the adversarial sampler optimization process. Next, we delve into the construction of the histogram error distribution.

\noindent \textbf{Meta Sampling  -- Histogram Error Distribution.} First, refer to reference~\cite{geman1992neural}, we give a formal definition of the histogram error distribution as follows:
\begin{equation}\label{eq-9}
\small
\hat E_{{D}}^i = \frac{{|\{ (x,y)|\frac{{i - 1}}{b} \leqslant {\ell }(f_s(\theta ,x) - y) < \frac{i}{b},(x,y) \in {D}\} |}}{{|{D}|}},
\end{equation}
where $1 \leqslant i \leqslant b$ and $\hat E_{{D}} \in {\mathbb{R}^b}$ is a vector that denotes the error distribution $E_{{D}} $ approximated by the histogram on dataset $D$, $\hat E_{{D}}^i$ denotes the $i$-th component of vector $\hat E_{{D}}$, $f_s$ is the surrogate model, and $b$ is the number of bins in the histogram. Gvien tranining set $D_{t}$ and validation set $D_v$, we have the meta-state:
   \vspace{-0.2cm}
\begin{equation}\label{eq-10}
    s=\left[\widehat{E}_{{D}_{t}}: \widehat{E}_{{D}_{v}}\right] \in \mathbb{R}^{2 b} .
    \vspace{-0.3cm}
\end{equation}
In practice, the histogram error distribution $\hat E_{{D}}$ reflects the fit of a given classifier to the dataset $D$. By adjusting $b$ in a fine-grained manner, it captures the distribution information of hard and easy samples, enriching the adversarial sampling process with more generalization error information. Moreover, in conjunction with Defi. \ref{defi-1}, the meta-state provides the meta-sampler with insights into the bias/variance of the current local model, thus supporting its decision-making. \textit{However, directly maximizing the generalization error solely based on this meta-state is impractical.} Sampling from large datasets for sample-level decisions incurs exponential complexity and time consumption. Additionally, blindly maximizing training and validation errors can be easily detected by an experienced defender.

\noindent \textbf{Meta Sampling -- Gaussian Sampling.} To make updating the meta-state more efficient, we leverage a \emph{Gaussian function trick} to simplify the meta-sampling process and the sampler itself, reducing the sampling complexity from $\mathcal{O}(|D|)$ to $\mathcal{O}(1)$. Specifically, we aim to construct a mapping such that an adversarial meta-sampler $\xi ( \cdot )$ outputs into a scalar interval $\mu  \in [0,1]$ for a given input meta-state $s$, \ie, $\mu  \sim \xi (\mu |s)$. In doing so, we apply a Gaussian function ${g_{\mu ,\sigma }}(x)$ to the classification error of each instance to determine its sampling weights, where ${g_{\mu ,\sigma }}(x)$ is defined as:
$g_{\mu, \sigma}(x)=\frac{1}{\sigma \sqrt{2 \pi}} e^{-\frac{1}{2}\left(\frac{x-\mu}{\sigma}\right)^2} ,$ where $e$ is the Euler’s number and $\sigma$ is a hyperparameter. Therefore, for the i-th sample, its sampled weight ${w_i} = \frac{{{g_{\mu ,\sigma }}\left( {\ell (f_s({x_i}|D),{y_i}} \right)}}{{\sum\limits_{\left( {{x_j},{y_j}} \right) \in D} {{g_{\mu ,\sigma }}} \left( {\left( {\ell (f({x_j}|D),{y_j}} \right)} \right)}}$.

\noindent \textbf{Meta Sampling Strategy -- Meta Training.} In this section, we present the training process of the adversarial meta-sampler and its strategy to maximize generalization error. The adversarial meta-sampler is designed to iteratively and adaptively select adversarial training batches, aiming to maximize the generalization error of the local model. As mentioned earlier, the sampler takes the current meta-state $s$ as input and outputs Gaussian function parameters to determine the sampling probability for each sample. By learning and adjusting its selection strategy through the state($s$)--action($\mu$)--state(new $s$) interaction, the adversarial meta-sampler seeks to maximize the generalization error. Consequently, the non-convex and nonlinear nature of the objective function is naturally addressed through Reinforcement Learning (RL)~\cite{haarnoja2018soft}. We consider the generalization error maximization process as an environment (ENV) in the RL setting. Therefore, in the RL setting, we treat the above optimization process as a Markov decision process (MDP), where the MDP can be defined by the tuple $(\mathcal{S},\Omega ,P_a,R_a)$. The definition of parameters in this tuple can be seen as follows: $\mathcal{S}$ is a set of state $s$ called state space, $\Omega$ is a set of actions $a$ called the action space (where $a_t \in \Omega \in [0,1]$ is continuous), $P_a\left(s, s^{\prime}\right)=\operatorname{Pr}\left(s_{t+1}=s^{\prime} \mid s_t=s, a_t=a\right)$ is the probability that action $a_t$ in state $s_t$ at time $t$ will lead to state $s_{t+1}$ at time $t+1$, and $R_a(s,s^{\prime})$ is the immediate reward received after transitioning from state $s$ to state $s^{\prime}$, due to action $a$. More specifically, at each environment step, ENV provides the meta-state ${s_t} = [{\hat E_{{D_{t}}}}:{\hat E_{{D_{v}}}}]$, and then the action $a_t$ is selected by ${a_t} \sim \xi ({\mu _t}|{s_t})$. The new state $s_{t+1}$ will be sampled $w.r.t{\text{ }}{s_{t + 1}} \sim {P_a}(s,{s^\prime })$ in the next round of local training. To maximize the generalization error (\ie, high validation error) while maintaining good learning behavior (\ie, low training error), we design a generalization error quantization function $G(\theta ,D) = \sum\limits_{i = 1}^{|D|} {\ell (f({x_i}|D),{y_i})} $, where $G(\theta ,D) = \sum\limits_{i = 1}^{|D|} {\ell ({f_\theta }({x_i}|D),{y_i})} $ is the sum of losses on the validation set $D_{v}$, then the reward $r$ can be defined as the difference in generalization performance of the surrogate model before and after the update, \ie, ${r_t} = \sum\limits_{i = 1}^{|{D_v}|} {\ell (f_k^{t + 1}({x_i} \in {D_v}),{y_i})}  - \sum\limits_{i = 1}^{|{D_{v|}}} {\ell (f_k^t({x_i} \in {D_v}),{y_i})}.$ Therefore, the above optimization objective (\ie, cumulative reward) of the adversarial meta-sampler is to maximize the generalization error of the surrogate model. To achieve this goal, we utilize the \emph{Soft Actor-Critic (SAC)} theory in~\cite{haarnoja2018soft}, to optimize our meta-sampler $\xi ( \cdot )$. The overview of the proposed attack algorithms are shown in Algo. \ref{alg-1}--\ref{algo-3} (refer to Appendix \ref{alg:attackalgo}).

\section{Theoretical Analysis}\label{sec-5}

\subsection{Error Upper Bound Analysis for BadSampler}\label{sec-5-1}
In analyzing the error lower bound of the proposed attack, it is crucial to consider that this bound can be influenced by factors such as the experimental environment, communication status, and network delay. As a consequence, the practical significance of conducting such an analysis is limited. Therefore, while understanding the theoretical lower bound is informative, it is essential to interpret the results within the context of specific experimental conditions and practical considerations. Thus, this paper provides a detailed error upper bound analysis. For simplicity, we make the required assumptions as follows. 

\begin{assumption} \label{assum-1}
(Bounded Gradients). For any model parameter $\omega$ and in the sequence $[{\omega _0},{\omega _1}, \ldots ,{\omega _t}, \ldots ]$, the norm of the gradient at every sample is bounded by a constant $\varepsilon_0^2$, i.e., $\forall \omega $, we have: $||\nabla {F}(\omega )||^2 \leqslant {\varepsilon_0^2}$.
\end{assumption}

\begin{assumption} \label{assum-3}
(Bounded BadSampler's Gradients). For any model parameter $\omega$ and in the sequence $[{\omega _0}^b,{\omega _1}^b, \ldots ,{\omega _t}^b, \ldots ]$, the norm of the bad batch's gradient at every sample is bounded by a constant $\varepsilon_b^2$, i.e., $\forall \omega^b $, we have: $||\nabla {F}(\omega^b )||^2 \leqslant {\varepsilon_b^2}$.
\end{assumption}

\begin{assumption} \label{assum-2}
(Unbiased Estimation Sampling). For any training epoch and any model parameter $\omega$, given an unbiased sampling, the expectation of the batch gradient matches the true gradient, \ie, $\forall \mathbf{b} \sim \mathcal{B}$ and $(x,y) \sim D$, we have: $\mathbb{E}[\nabla \ell _{{i_t}}({\omega})] = \frac{1}{N}\sum\limits_{i = 1}^N \nabla  \ell _i({\omega }) = \nabla {F}({\omega }),$ and ${\mathbb{P}({i_t} = i)}=\frac{1}{N}$.
\end{assumption}

To analyze the error upper bound of \emph{\textsf{BadSampler}} attack, we give Lemma \ref{lemm-1} and Lemma \ref{lemm-2} as follows:
\begin{lemma}\label{lemm-1}
If Assumptions \ref{assum-1} and \ref{assum-2} hold, the expectation of the stochastic second-order correction term, \ie, the expectation of the bias, is formally expressed as follows:
\begin{equation}\label{eq-15}
\small
\begin{aligned}[c]
  \mathbb{E}[\xi (\omega )] &= \frac{{{N^2}}}{4}\nabla (||\nabla F(\omega )|{|^2} - \frac{1}{{{N^2}}}\sum\nolimits_{j = 0}^{N - 1} {||{\ell _j}(\omega )|{|^2})}.
\end{aligned} 
\end{equation}
\end{lemma}

\begin{proof}
Full proofs can be found in the Appendix \ref{proof}.
\end{proof} 

\begin{lemma}\label{lemm-2}
If Assumptions \ref{assum-1}, \ref{assum-2} and Lemma \ref{lemm-1} hold, the expectation of the local model parameter term is formally expressed as follows:
\begin{equation}\label{eq-16}
\centering
\begin{aligned}
  \mathbb{E}[{\omega _t}]& = {\omega _0} - N\eta \nabla F({\omega _0}) 
   + \frac{{{N^2}{\eta ^2}}}{4}\nabla (||\nabla F({\omega _0})|{|^2} \hfill \\&- \frac{1}{{{N^2}}}\sum\nolimits_{j = 0}^{N - 1} {||} {\ell _j}({\omega _0})|{|^2}) + O({N^3}{\eta ^3}). \hfill \\ 
\end{aligned} 
\end{equation}
\end{lemma}

\begin{proof}
Full proofs can be found in the Appendix \ref{proof}.
\end{proof} 

Therefore, the following Theorem \ref{theo-1} gives the bias upper bound for our \emph{\textsf{BadSampler}} attack:

\begin{theorem}\label{theo-1}
We use $\omega^*$ to denote the model converged without any attack, and $\omega^b$ to denote the final model obtained under the BadBacth attack. If Assumptions \ref{assum-1}--\ref{assum-2} and Lemma \ref{lemm-1}--\ref{lemm-2} hold, the expectation of our attack's error is formally expressed as follows:
\begin{equation}\label{eq-17}
\small
\mathbb{E}[||\nabla F({\omega ^*}) - \nabla F({\omega ^b})|{|^2}] \leqslant B{(1 - \frac{1}{N})^2}\varepsilon _0^2 - \frac{{B - 1}}{{NB}}\varepsilon _b^2.
\vspace{-0.2cm}
\end{equation}
\end{theorem}
\begin{proof}
Full proofs can be found in the Appendix \ref{proof}.
\end{proof} 

\subsection{Complexity Analysis}\label{sec-5-2}
We give a formal complexity analysis of the proposed attacks. We denote the feature dimensions of the input sample as $d$, the number of training batches as $N$, the size of the training batch as $B$, and the total computation per sample (depending on the gradient computation formula) as $C$.

\noindent \textbf{Complexity of Top-$\kappa$ Sampling.} Let $\mathcal{O}(N\log \kappa )$ be the time complexity of the Top-$\kappa$ operation and let $\mathcal{O}(dBC)$ (or $\mathcal{O}(dC)$) be the time complexity of computing the gradient of a single batch (or sample), then the total time complexity of the \emph{\textsf{BadSampler}} attack to complete the sorting and compute the difficulty is $\mathcal{O}(N\log \kappa ) + \mathcal{O}(dNBC)$. For the adversary, we assume that it manipulates $c$ compromised clients, then the overall time complexity of the proposed attack is $\mathcal{O}(cN\log c(\log \kappa  + dBC))$.

\noindent \textbf{Complexity of Meta Sampling.} Compared with the Top-$\kappa$ sampling attack strategy, this attack strategy only brings additional overhead of performing meta-training. Therefore, we next give the required complexity for meta-training. Likewise, we assume that the cost of the meta-sampler to perform a single gradient update step is $C_{update}$ (depending on the gradient computation formula). In our attack implementation, following the SAC optimization strategy, we need to perform $n_{random}$ actions before updating the meta-sampler, and then need $n_{update}$ steps to collect online transitions and perform gradient updates for $\xi ( \cdot )$. Therefore, the overall cost of meta-training can be expressed as $O(({n_{random}} + {n_{update}})NBC) + O({n_{update}}{C_{update}})$. For the adversary, we assume that it manipulates $c$ compromised clients, then the overall time complexity of the proposed attack is $\mathcal{O}(cN\log c(\log \kappa  + dBC)) +O(c\log c(({n_{random}} + {n_{update}})NBC + {n_{update}}{C_{update}}))$.

\section{Experiments}\label{sec-6}
\subsection{Experiment Setup}\label{sec-6-1}
To evaluate the performance of our attacks, we conduct extensive experiments on two benchmark datasets. All designs are developed using Python 3.7 and PyTorch 1.7 and evaluated on a server with an NVIDIA GeForce RTX2080 Ti GPU and an Intel Xeon Silver 4210 CPU.

\noindent \textbf{Datasets.} We adopt two image datasets for evaluations, \ie, \textsf{Fashion-MNIST} (F-MNIST)~\cite{xiao2017fashion} and \textsf{CIFAR-10}~\cite{krizhevsky2009learning}. The datasets cover different attributes, dimensions, and numbers of categories, allowing us to explore the poisoning effectiveness of \emph{\textsf{BadSampler}}. To simulate the \emph{IID} setting of FL, we evenly distribute the two training datasets to all clients, and to simulate \emph{non-IID} setting of FL, we follow~\cite{mcmahan2017communication} to distribute the two training datasets to all clients (see Appendix~\ref{appd-3}).

\noindent \textbf{Models.} In this experiment, we validate our \emph{\textsf{BadSampler}} attack in both convex and non-convex settings. For convex problems, we use the \emph{Logistic Regression (LR)} model as our local model; for non-convex problems, we use a simple \emph{CNN} model, \ie, CNN with 2 convolutional layers followed by 1 fully connected layer and the \emph{ResNet-18} model. Furthermore, we use a common CNN model, \ie, \emph{LeNet-5} model as our surrogate model.

\noindent \textbf{Hyperparameters.} In our attacks, we consider the cross-device FL scenario and we set the number of clients $K=100$ and the proportion of client participation $q=10\%$. Unless otherwise mentioned, we set local training epoch ${E_{local}}=5$, the learning rate $\eta  = 0.001$, the total number of training rounds $T=250$, the Gaussian function parameter $\sigma = 0.2$, the meta-state size is 10 (\ie, $b=5$), the proportion of the compromised clients $M=\{5\%,10\%\}$, the mini-batch size $B \in \{8,16,32,64\}$ and constant $\kappa \in \{1, 2, 4, 8\}$. Table \ref{tab:practical_ranges} summarizes our practical parameters/settings for comparing the previous work.

\noindent \textbf{Attacks for Comparison.} We compare the following attacks with \emph{\textsf{BadSampler}}. \textit{Label Flipping Attack (LFA)~\cite{ref-78,tolpegin2020data}:} a classic data poisoning attack where the adversary modifies one class of true label of the local dataset to another class. \textit{Adversarial Attack (AA)~\cite{ref-14}:} a recent data poisoning attack, where the adversary accesses a local training dataset to generate an adversarial dataset $D_{adv}$ to poison the global model. \textit{Gaussian Attack (GA)~\cite{ref-13,ref-72}:} a classic model poisoning attack, where the adversary uploads its model update $\omega _t^k$ from a Gaussian distribution with mean $\frac{1}{{k - m}}\sum\nolimits_{k \notin {M}} {\omega _t^k} $ and variance 10. \textit{Zero-update Attack (ZA)~\cite{lin2019byzantine,ref-76}:} a powerful local model poisoning attack, where the adversary takes control of the compromised clients and lets them send zero-sum model updates, \ie, ${\omega _p} = \sum\nolimits_{k \in M} {\omega _t^k}  = 0$. \textit{Local Model Poisoning Attack (LMPA)~\cite{ref-13}:} a state-of-the-art local model poisoning attack, in which the adversary controls a small set of compromised clients to jointly optimize a well-designed model-directed deviation objective function to poison the FL model. \textit{OBLIVION~\cite{zhang2023oblivion}: a model poisoning attack that uses modified model weights to achieve catastrophic forgetting.} \textit{Data Ordering Attack (ODA)~\cite{shumailov2021manipulating}: a data poisoning attack that uses modifying the order of training data to achieve catastrophic forgetting.} As mentioned above, we chose both the classic data poisoning attack and the model poisoning attack to compare the proposed attacks fairly.


%


\noindent \textbf{Defenses.} To further validate the performance of \emph{\textsf{BadSampler}}, we evaluate attacks under three \textit{anomaly-detection-based defenses}, \ie, \textsf{PCA}-based defense~\cite{tolpegin2020data}, \textsf{FoolsGold}~\cite{ref-78}, and FLDetector~\cite{zhang2022fldetector}, and three \textit{Byzantine robust AGRs}, \ie, \textsf{Trimmed Mean}~\cite{yin2018byzantine},  \textsf{Krum}~\cite{blanchard2017machine}, and \textsf{FLTrust}~\cite{caofltrust}. \textit{Note that here we do not consider the verification set defense scheme because the verification set is already used in \textsf{FLTrust.}}

%

\noindent \textbf{Attack Impact Metric.} We denote by $Acc_{test}^*$ the maximum accuracy that the global model converges without any attack. We use $Acc_{test}^p$ to denote the maximum accuracy the global model can achieve under a given attack. Here, we define attack impact $\Delta $ as the reduction in global model accuracy due to the attack, thus for a given attack, \ie, $\Delta  = Ac{c_{test}}^* - Ac{c_{test}}^p$.

\subsection{Evaluation}\label{sec-6-2}

\begin{table}[!t]
\caption{Attack impact in the production federated learning settings.}
\label{tab-5}
\resizebox{\columnwidth}{!}{%
\begin{tabular}{@{}lccc||ccc||ccc@{}}
\toprule
\toprule
\multicolumn{1}{c}{\multirow{2}{*}{Attacks}} & \multicolumn{3}{c}{LR (F-MNIST)} & \multicolumn{3}{c}{CNN (CIFAR-10)} & \multicolumn{3}{c}{ResNet-18 (CIFAR-10)} \\ \cmidrule(l){2-10} 
\multicolumn{1}{c}{} & \begin{tabular}[c]{@{}c@{}}Train \\ acc (\%)\end{tabular} & \begin{tabular}[c]{@{}c@{}}Test\\ acc (\%)\end{tabular} & $\Delta \uparrow$& \begin{tabular}[c]{@{}c@{}}Train \\ acc (\%)\end{tabular} & \begin{tabular}[c]{@{}c@{}}Test\\ acc (\%)\end{tabular} & $\Delta \uparrow$ & \begin{tabular}[c]{@{}c@{}}Train \\ acc (\%)\end{tabular} & \begin{tabular}[c]{@{}c@{}}Test\\ acc (\%)\end{tabular} & $\Delta \uparrow$ \\ \cmidrule(r){1-10}
None &86.03  &85.08  &0  &63.73  &59.10  &0  &92.34  &67.11  &0  \\

LFA &85.41  &83.94  &1.14  &60.65  &57.21  &1.89  &91.88  &65.58  &1.53  \\

AA &83.06  &82.41  &2.67  &58.96  &55.07  &4.03  &88.36  &63.88  &3.29  \\

GA &82.63  &79.46  &6.34  &59.24  &53.96  &5.44  &85.63  &60.85  &6.26  \\

ZA &86.01  &84.99  &0.09  &62.97  &58.41  &0.69  &91.06  &65.32  &1.79  \\

LMPA &79.31  &75.64  &9.44  &57.96  &52.11  &6.99  &84.69  &61.64  &5.47  \\ 

DOA &85.42  &81.76  &3.32  &62.88  &55.41  &3.69  &90.86  &65.42  &1.69  \\

OBLIVION &80.42  &77.83  &7.25  &60.85  &53.32  &5.78  &83.55  &60.67  &6.44 \\ \bottomrule

\underline{\textit{Ours}}\\
  Top-$\kappa$  &76.07  &75.87  &\textbf{9.21}  &43.60  &40.92  &\textbf{18.18}  &72.73  &56.38  &\textbf{10.73}  \\
  
   Meta &80.78  &72.12  &\textbf{12.96}  &49.17  &39.80  &\textbf{19.30}  &87.92  &56.74  &\textbf{10.37}  \\ \bottomrule \bottomrule
\end{tabular}%
\vspace{-0.4cm}
}
\end{table}

\subsubsection{Attack Performance under Practical Settings} \label{sec-6-2-1} 
In the first evaluation, we evaluate the effectiveness of our proposed \textsf{BadSampler} attack under a realistic setting, considering practical parameter ranges (refer to Table \ref{tab:practical_ranges}) in FL. We set the batch size $B=32$, the number of candidate adversarial training batches $\kappa = 2$, and use FedAvg~\cite{mcmahan2017communication} as AGR. For classification tasks on the \textsf{Fashion-MNIST} dataset, we employ the LR model, and for the \textsf{CIFAR-10} dataset, we use the CNN model and ResNet-18 model. The adversary's capabilities are restricted; they do not have access to global model parameters and must carry out the attack within strict parameters. We employ the LeNet model as the surrogate model to compute sample difficulty. Table \ref{tab-5} demonstrates that despite these limitations, our designed attack scheme outperforms selected baselines, particularly the adversarial meta-sampling attack, where our attack achieves a remarkable performance of 19.3\%. The key to our success lies in adaptively sampling and feeding the model hard samples during the learning process, which can destroy the model's generalization performance and cause the model to fall into catastrophic forgetting. 

\subsubsection{Attack Performance under Byzantine Robust Aggregation based Defenses} \label{sec-6-2-4} To validate the effectiveness of our \emph{\textsf{BadSampler}} attack, we evaluate its performance against \textsf{Trimmed Mean}~\cite{yin2018byzantine}, \textsf{Krum}~\cite{blanchard2017machine}, and \textsf{FLTrust}~\cite{caofltrust}, which are server-side Byzantine robust aggregation-based defenses. These defenses have previously demonstrated robustness against existing DPAs and MPAs. In this experiment, we conduct comparative evaluations using the CNN model on the \textsf{CIFAR-10} dataset. Table \ref{tab-10} presents the attack impact $\Delta$ of our attack against the three defenses on \textsf{CIFAR-10}. Notably, our attack successfully leads to a substantial drop in global model accuracy, even in the presence of strong defense measures. For instance, when confronted with the state-of-the-art defense \textsf{FLTrust}, our attack reduces the convergence accuracy of the CNN model by $8.98\%$. Byzantine robust aggregation defenses are effective against attacks involving malicious model updates due to their reliance on specific aggregation rules. However, these defenses face challenges in countering our attack, as they disrupt the generalization ability of the model, making it difficult for the aggregation rules to enhance model generalization.

\subsubsection{Attack Performance under Anomaly Detection Based Defenses} \label{sec-6-2-3} 
To evaluate the attack performance of \emph{\textsf{BadSampler}} and baselines under Byzantine-robust FL with anomaly detection defenses, we use the \textsf{PCA}-based method~\cite{tolpegin2020data}, \textsf{FoolsGold}~\cite{ref-78}, and FLDetector~\cite{zhang2022fldetector} defenses against the proposed attacks. Specifically, we evaluate the attack impact $\Delta$ of \emph{\textsf{BadSampler}} on CNN models on \textsf{CIFAR-10} dataset. We fix the local training epoch $E_{local}=5$, the batch size $B=32$, and the number of candidate adversarial training batches $\kappa = 2$. Table \ref{tab-9} shows that our attacks still effectively increase the classification error of the aggregated model. For example, under the \textsf{PCA}-based defense, our attack using the CNN model as a classifier can cause a 12.85\% accuracy loss on the \textsf{CIFAR-10} dataset. The main reason is that anomaly detection-based defenses are less effective when the compromised client is not behaving maliciously. In addition, these defenses will also bring accuracy loss to the model due to the dimensionality reduction operation involved.

\begin{table}[!t]
\caption{Attack impact under different Byzantine robust AGRs.}
\label{tab-10}
\resizebox{\columnwidth}{!}{%
\begin{tabular}{@{}lccc||ccc||ccc@{}}
\toprule
\toprule
\multicolumn{1}{c}{\multirow{2}{*}{Attacks}} & \multicolumn{3}{c}{Trimmed Mean} & \multicolumn{3}{c}{Krum} & \multicolumn{3}{c}{FLTrust} \\ \cmidrule(l){2-10} 
\multicolumn{1}{c}{} & \begin{tabular}[c]{@{}c@{}}Train \\ acc (\%)\end{tabular} & \begin{tabular}[c]{@{}c@{}}Test\\ acc (\%)\end{tabular} & $\Delta \uparrow$& \begin{tabular}[c]{@{}c@{}}Train \\ acc (\%)\end{tabular} & \begin{tabular}[c]{@{}c@{}}Test\\ acc (\%)\end{tabular} & $\Delta \uparrow$ & \begin{tabular}[c]{@{}c@{}}Train \\ acc (\%)\end{tabular} & \begin{tabular}[c]{@{}c@{}}Test\\ acc (\%)\end{tabular} & $\Delta \uparrow$ \\ \cmidrule(r){1-10}
None &63.73  &59.10  &0  &63.73  &59.10  &0  &63.73  &59.10    &0  \\
LFA &63.12  &58.89  &0.21  &63.54  &58.99  &0.11  &63.50  &59.00  &0.10  \\

AA &63.01  &58.68  &0.42  &63.25  &59.01  &0.09  &63.64  &59.07  &0.03  \\

GA &63.45  &58.24  &0.86  &63.32  &58.71  &0.39  &63.65  &59.04  &0.06  \\

ZA &63.70  &59.10  &0  &63.66 &59.07  &0.03 &63.64  &59.10  &0  \\

LMPA &59.24  &54.97  &4.13  &59.98  &55.21  &3.89  &61.24  &56.65  &2.45  \\ 

DOA &61.57  &56.21  &2.89  &61.89  &56.42  &2.68  &62.10  &57.43  &1.67  \\

OBLIVION &57.32  &53.96  &5.14  &57.54  &52.32  &5.78  &57.41  &56.17  &2.93   \\ \bottomrule

\underline{\textit{Ours}}\\
  Top-$\kappa$  &55.65  &52.24  &\textbf{7.86}  &54.60  &50.12  &\textbf{8.98}  &55.70  &51.21  &\textbf{7.89}  \\
  
   Meta &54.24  &49.04  &\textbf{10.06}  &54.27  &50.09  &\textbf{9.09}  &55.04  &50.12  &\textbf{8.98}  \\ \bottomrule \bottomrule
\end{tabular}%
\vspace{-0.3cm}
}
\end{table}

\begin{table}[!t]
\caption{Attack impact under anomaly detection based defenses.}
\label{tab-9}
\resizebox{\columnwidth}{!}{%
\begin{tabular}{@{}lccc||ccc||ccc@{}}
\toprule
\toprule
\multicolumn{1}{c}{\multirow{2}{*}{Attacks}} & \multicolumn{3}{c}{PCA} & \multicolumn{3}{c}{FoolsGold} & \multicolumn{3}{c}{FLDetector} \\ \cmidrule(l){2-10} 
\multicolumn{1}{c}{} & \begin{tabular}[c]{@{}c@{}}Train \\ acc (\%)\end{tabular} & \begin{tabular}[c]{@{}c@{}}Test\\ acc (\%)\end{tabular} & $\Delta \uparrow$& \begin{tabular}[c]{@{}c@{}}Train \\ acc (\%)\end{tabular} & \begin{tabular}[c]{@{}c@{}}Test\\ acc (\%)\end{tabular} & $\Delta \uparrow$ & \begin{tabular}[c]{@{}c@{}}Train \\ acc (\%)\end{tabular} & \begin{tabular}[c]{@{}c@{}}Test\\ acc (\%)\end{tabular} & $\Delta \uparrow$ \\ \cmidrule(r){1-10}
None &63.73  &59.10  &0  &63.73  &59.10  &0  &63.73  &59.10    &0  \\

LFA &62.44  &57.90  &1.20  &62.14  &57.54  &1.56  &63.50  &58.63  &0.47  \\

AA &62.35  &57.68  &1.42  &62.78  &57.65  &1.45  &63.64  &59.07  &0.03  \\

GA &63.52  &58.36  &0.74  &63.57  &58.75  &0.35  &63.65  &59.05  &0.05  \\

ZA &63.70  &59.10  &0  &63.66 &59.07  &0.03 &63.64  &59.10  &0  \\

LMPA &56.11  &53.21  &5.89  &56.54  &54.12  &4.95  &62.25  &57.01  &2.09  \\ 

DOA &62.14  &56.87  &2.23  &62.64  &57.16  &1.94  &63.10  &57.54  &1.56  \\

OBLIVION &57.12  &53.32  &5.78  &57.21  &52.01  &7.09  &58.55  &53.17  &5.93   \\ \bottomrule

\underline{\textit{Ours}}\\
  Top-$\kappa$  &54.21  &49.29  &\textbf{9.81}  &54.77  &50.25  &\textbf{8.85}  &55.65  &52.22  &\textbf{6.88}  \\
  
   Meta &50.54  &46.25  &\textbf{12.85}  &51.98  &48.02  &\textbf{11.08}  &55.41  &50.65  &\textbf{8.45}  \\ \bottomrule \bottomrule
\end{tabular}%
}
\end{table}

\subsubsection{Attack Performance under Different Hyperparameters} \label{sec-6-2-5} We summarize the experimental results as follows:

\noindent \textbf{Impact of the Parameter $M$ in \emph{\textsf{BadSampler}} Attacks.} Table \ref{tab-3} shows the attack impact of different attack strategies as the percentage of compromised clients increases on the \textsf{CIFAR-10} dataset using the CNN model. Experimental results show that our attack significantly degrades the aggregation model's performance as the proportion of compromised clients increases. In particular, our attack can still effectively attack the aggregated model when the proportion of compromised clients is merely 5\%. Compared to other common data poisoning attacks, our attack is more robust and practical.

\noindent \textbf{Impact of the Parameter $\kappa$ in \emph{\textsf{BadSampler}} Attacks.} We explore the effect of the poisoned sample size on Top-$\kappa$ sampling attack performance. In this experiment, we change the poisoned sample size by adjusting the number of adversarial training batches $\kappa=\{1,2,4,8\}$. Table \ref{tab-3} shows that within a certain number range, the more adversarial training batches $\kappa$ there are, the better the attack performance is. For example, when $\kappa=4$, our \emph{\textsf{BadSampler}} attack performance does not change significantly compared to it when $\kappa=1$. The reason is that the error caused by our attack has an upper bound, that is, the number of adversarial training batches cannot be increased infinitely to enhance the attack performance.

\noindent \textbf{Impact of the Parameter $B$ in \emph{\textsf{BadSampler}} Attacks.} Here we show that if we fix the dataset size, the larger the batch size, the less the number of adversarial training batches we can choose. Table \ref{tab-3} reports the attack impact $\Delta$ when $B=\{16,32,64\}$. The experimental results show that the batch size gradually increases, and the impact of the attack gradually weakens, but it is still stronger than the latest poisoning attacks. Furthermore, we emphasize that small batch size are commonly used for training in resource-constrained scenarios \cite{mcmahan2017communication,li2020federated}. Therefore, our \emph{\textsf{BadSampler}} attack is practical.


\begin{table}[!t]
\centering
\caption{Attack impact under different hyperparameters.}
\label{tab-3}
\resizebox{\columnwidth}{!}{%
\begin{tabular}{@{}ccccccccccc@{}}
\toprule
\toprule
\multirow{3}{*}{Attacks} &
  \multicolumn{10}{c}{Hyperparameters} \\ \cmidrule(l){2-11} 
 &
  \multicolumn{2}{c|}{$M$} &
  \multicolumn{4}{c|}{$\kappa$} &
  \multicolumn{4}{c}{$B$} \\ \cmidrule(l){2-11} 
 &
  \multicolumn{1}{c}{5\%} &
  \multicolumn{1}{c|}{10\%} &
  \multicolumn{1}{c}{1} &
  \multicolumn{1}{c}{2} &
  \multicolumn{1}{c}{4} &
  \multicolumn{1}{c|}{8} &
  \multicolumn{1}{c}{8} &
  \multicolumn{1}{c}{16} &
  \multicolumn{1}{c}{32} &
  \multicolumn{1}{c}{64} \\ \midrule
\multicolumn{1}{c}{Top-$\kappa$ (\%)} &
  \multicolumn{1}{c}{7.3} &
  \multicolumn{1}{c|}{18.18} &
  \multicolumn{1}{c}{17.5} &
  \multicolumn{1}{c}{18.18} &
  \multicolumn{1}{c}{13.8} &
  \multicolumn{1}{c|}{10.2} &
  \multicolumn{1}{c}{8.3} &
  \multicolumn{1}{c}{9.8} &
  \multicolumn{1}{c}{18.18} &9.4
   \\ 
\multicolumn{1}{c}{Meta (\%)} &
  \multicolumn{1}{c}{8.6} &
  \multicolumn{1}{c|}{19.3} &
  \multicolumn{1}{c}{N/A} &
  \multicolumn{1}{c}{N/A} &
  \multicolumn{1}{c}{N/A} &
  \multicolumn{1}{c|}{N/A} &
  \multicolumn{1}{c}{12.6} &
  \multicolumn{1}{c}{13.8} &
  \multicolumn{1}{c}{19.3} &11.3
   \\ \bottomrule \bottomrule
\end{tabular}%
\vspace{-0.5cm}
}
\end{table}

\begin{table}[!t]
\caption{Attack impact under the non-IID setting.}
\label{tab-8}
\resizebox{\columnwidth}{!}{%
\begin{tabular}{@{}lccc||ccc||ccc@{}}
\toprule
\toprule
\multicolumn{1}{c}{\multirow{2}{*}{Attacks}} & \multicolumn{3}{c}{LR (F-MNIST)} & \multicolumn{3}{c}{CNN (CIFAR-10)} & \multicolumn{3}{c}{ResNet-18 (CIFAR-10)} \\ \cmidrule(l){2-10} 
\multicolumn{1}{c}{} & \begin{tabular}[c]{@{}c@{}}Train \\ acc (\%)\end{tabular} & \begin{tabular}[c]{@{}c@{}}Test\\ acc (\%)\end{tabular} & $\Delta \uparrow$& \begin{tabular}[c]{@{}c@{}}Train \\ acc (\%)\end{tabular} & \begin{tabular}[c]{@{}c@{}}Test\\ acc (\%)\end{tabular} & $\Delta \uparrow$ & \begin{tabular}[c]{@{}c@{}}Train \\ acc (\%)\end{tabular} & \begin{tabular}[c]{@{}c@{}}Test\\ acc (\%)\end{tabular} & $\Delta \uparrow$ \\ \cmidrule(r){1-10}
None &82.51  &81.99  &0  &52.34  &47.45  &0  &36.67  &34.15  &0  \\
Top-$\kappa$ (Ours) &71.88  &71.29  &\textbf{10.7}  &32.99  &31.20  &\textbf{16.25}  &22.82  &22.34 &\textbf{12.81}  \\
Meta (Ours) &80.27  &68.84  &\textbf{13.15}  &48.36  &35.47  &\textbf{11.98}  &31.91  &20.29  &\textbf{13.86}  \\ \bottomrule \bottomrule
\end{tabular}%
\vspace{-0.5cm}
}
\end{table}


\subsubsection{Attack Performance under Non-IID Setting} \label{sec-6-2-6} Last, we evaluate the attack performance under the non-IID data setting ($\alpha=0.5$). In this experiment, we fix the local training epoch $E_{local}=5$, the batch size $B=32$, and the number of adversarial training batches $\kappa = 2$. We use the LR model for classification tasks on the \textsf{FMNIST} dataset and the CNN model and ResNet-18 model for classification tasks on the \textsf{CIFAR-10} dataset. Table \ref{tab-8} shows that our attacks are still effective in the non-IID setting.

\subsubsection{Effectiveness of \textsf{BadSampler}}
Here, we will demonstrate why \emph{\textsf{BadSampler}} is effective from the perspective of training behavior, generalization ability evaluation, and generalization ability visualization. To this end, we use the following three metrics~\cite{li2018visualizing,mendieta2022local}: the gradient cosine distance: $\mathcal{D}_{\cos }$, the difference in Hessian norm ($H_N$), and the difference in Hessian direction ($H_D$). We conducted five sets of experiments and recorded the mean value of the experimental results of these five sets of experiments in Table \ref{tab-6}. The results show that the training behavior of our attack is not very different from that of benign clients, especially the meta-sampling attack strategy. This means that it is difficult for sophisticated defenders to quickly determine which clients are malicious, which brings new challenges to current advanced defenses.
Furthermore, we use the parametric loss landscape~\cite{li2018visualizing} visualization with Hessian feature eigenvectors (${\lambda _{ma{x_x}}}$ and ${\lambda _{ma{x_y}}}$) for the resulting global model before and after the attack to show the impact of our attack on the model generalization ability. From Fig. \ref{fig:my_label}, we can clearly see that the loss landscape of the poisoned model changes more drastically and its optimization route is not smooth, even the route optimized to the local optimum is extremely uneven. This means that the generalization ability of the poisoned model is greatly damaged. On the contrary, in the absence of attacks, the loss landscape of the global model changes steadily and its optimization route is smooth and flat. Please find more details in the Appendix \ref{appd-4}.

\subsubsection{Overhead of \textsf{BadSampler}} We recorded the running time of the attacks over ten rounds to underscore the low attack-effort characteristic of our approach. Table \ref{tab-7} reveals that the proposed attack's running time is lower than LMPA and comparable to DOAs, which signifies the efficiency of our attack. This efficiency is attributed to the utilization of Top-$\kappa$ and RL techniques, which enable us to circumvent complex optimization processes. Additional experimental results can be found in Appendix \ref{appd-3}.

\begin{table}[!t]
\centering
\caption{Numerical results (Group 1: Benign and Benign; Group 2: Benign and Malicious).}
\label{tab-6}
\resizebox{\columnwidth}{!}{%
\begin{tabular}{@{}ccccccc@{}}
\toprule
\toprule
\multirow{2}{*}{Attacks} & \multicolumn{2}{c}{ $\mathcal{D}_{cos}$} & \multicolumn{2}{c}{$H_N$} & \multicolumn{2}{c}{$H_D$} \\ \cmidrule(l){2-7} 
 &
  \multicolumn{1}{c|}{Group 1} &
  \multicolumn{1}{c|}{Group 2} &
  \multicolumn{1}{c|}{Group 1} &
  \multicolumn{1}{c|}{Group 2} &
  \multicolumn{1}{c|}{Group 1} &
  \multicolumn{1}{c}{Group 2} \\ \midrule
Top-$\kappa$            & 1.09                & 0.73                & 8456               & 13256              & 0.98                       & 0.67                      \\

Meta             & 1.12                & 0.89                & 9687               & 15428              & 0.97                      & 0.96                      \\ \bottomrule \bottomrule
\end{tabular}%
\vspace{-0.5cm}
}
\end{table}

\begin{table}[!t]
\centering
\footnotesize
\caption{Numerical result of running time.}
\label{tab-7}
\resizebox{\columnwidth}{!}{%
\begin{tabular}{@{}l|ccccc@{}}
\toprule
\toprule
Attacks          & OBLIVION  & LMPA   & DOA   & Top-$\kappa$ & Meta   \\ \midrule
Running Time (s) & 185.24 & 246.32 & 154.54 &  \textbf{86.22} & \textbf{156.48} \\ \bottomrule \bottomrule
\end{tabular}%
\vspace{-0.5cm}
}
\end{table}

\section{Conclusions}
In this work, we introduced \emph{\textsf{BadSampler}}, a clean-label data poisoning attack algorithm targeting Byzantine-robust FL. Such attacks are designed to maximize the generalization error of the model by performing adversarial sampling on clean-label data while maintaining good training behavior during training. Furthermore, we emphasize that the attack is learnable and adaptive so that it can be deployed to real FL scenarios and remains effective within strict parameter constraints. The proposed \emph{\textsf{BadSampler}} attack undergoes theoretical analysis and experimental evaluation, demonstrating its effectiveness on moderate-scale public datasets with various parameter ranges, defense strategies, and FL deployment scenarios, highlighting its significant threat to real FL systems. 


\bibliographystyle{ACM-Reference-Format}
\bibliography{sample-base}
\clearpage
\appendix
\section{Production Federated Learning}\label{sec-3-1}
\noindent \textbf{Production FL Settings.}  Prior works~~\cite{ref-13,ref-15,ref-14,yi2023ua,wu2022fedattack} make several unrealistic assumptions in FL poisoning attacks in terms of the total number of clients, the percentage of compromised clients, and the knowledge held by the adversary. 
This means that the current literature is still far from practice.
Our approach focuses on creating a new poisoning attack in this realistic environment to close this gap.
Briefly, the difference between the production FL environment and the FL environment in the literature is the range of practical FL parameters. 
Following the work~\cite{shejwalkar2021back}, we summarize the differences between the practical FL parameters (called production FL parameters~\cite{shejwalkar2021back}) and the ones used in existing attacks in Table \ref{tab:practical_ranges}. 
Nonetheless, in this paper, our attack still needs to effectively attack the FL system in such a setting. 
Notably, to confirm the efficacy of our approach, our experimental setup adheres to every practical parameter setting in Table \ref{tab:practical_ranges}. 

\begin{table}[!t]
\caption{Summary of practical parameters/settings for production FL~\cite{shejwalkar2021back}. Note that we use green \textit{vs} red to distinguish whether parameters/settings are practical or not.} \label{tab:practical_ranges}
\centering
\resizebox{0.45\textwidth}{!}{%
\begin{tabular} {|c|c|c|}
  \hline
\bf \tabincell{c}{Parameters/Settings} & \bf  What we argue to be practical  & \bf  \tabincell{c}{Used in previous\\ \emph{data poisoning} works} \\ \hline
  
  \multirow{2}{*}{\tabincell{c}{FL type + Attack type}} & \multirow{2}{*}{\tabincell{c}{Cross-silo + DPAs\\ Cross-device + DPAs}} & \cellcolor{red!25} \\ 
  & & \multirow{-2}{*}{\cellcolor{red!25}{Cross-silo + MPAs}} \\ \hline
  
  \multirow{2}{*}{\tabincell{c}{Total number of FL\\ clients, $K$}} & \multirow{2}{*}{\tabincell{c}{Order of  $[10^1,10^{10}]$ for cross-device\\ $[2, 10]$ for cross-silo}} & \cellcolor{red!25} \\ 
  &  & \multirow{-2}{*}{\cellcolor{red!25}[50, 100]} \\ \hline
  
  \multirow{2}{*}{\tabincell{c}{Number of clients\\ chosen per round, $k$}} & \multirow{2}{*}{\tabincell{c}{Small fraction $q$ of $K$ for cross-device\\ All for cross-silo}} & \cellcolor{red!25} \\
  & & \multirow{-2}{*}{\cellcolor{red!25}All} \\ \hline
  
  \tabincell{c}{\% of compromised\\ clients, $M$} & \tabincell{c}{$M\leq$10\% for DPAs} & \cellcolor{red!25}$[20, 50]\%$  \\ \hline
  
  \tabincell{c}{Average size of benign\\ clients' data, $|D|_\mathsf{avg}$} & \tabincell{c}{$[50, 1000]$ for cross-device} & \tabincell{c}{Not studied for cross-device\\$[50,1000]$ for cross-silo} \\ \hline
  
  \tabincell{c}{Maximum size of\\ local poisoning data} & \tabincell{c}{Up to $100\times |D|_k$ for DPAs} & \cellcolor{green!25} $\sim |D|_k$ \\
  \hline
\end{tabular}
\vspace{-0.5cm}
}
\end{table}
\begin{table}[!t]
\caption{Attack impact with different surrogate models.}
\label{tab-11}
\resizebox{\columnwidth}{!}{%
\begin{tabular}{@{}lccc||ccc||ccc@{}}
\toprule
\toprule
\multicolumn{1}{c}{\multirow{2}{*}{Attacks}} & \multicolumn{3}{c}{LeNet-5} & \multicolumn{3}{c}{DenseNet} & \multicolumn{3}{c}{AlexNet} \\ \cmidrule(l){2-10} 
\multicolumn{1}{c}{} & \begin{tabular}[c]{@{}c@{}}Train \\ acc (\%)\end{tabular} & \begin{tabular}[c]{@{}c@{}}Test\\ acc (\%)\end{tabular} & $\Delta \uparrow$& \begin{tabular}[c]{@{}c@{}}Train \\ acc (\%)\end{tabular} & \begin{tabular}[c]{@{}c@{}}Test\\ acc (\%)\end{tabular} & $\Delta \uparrow$ & \begin{tabular}[c]{@{}c@{}}Train \\ acc (\%)\end{tabular} & \begin{tabular}[c]{@{}c@{}}Test\\ acc (\%)\end{tabular} & $\Delta \uparrow$ \\ \cmidrule(r){1-10}
None &63.73  &59.10  &0  &63.73  &59.10  &0  &63.73  &59.10  &0  \\
Top-$\kappa$ (Ours) &43.60  &40.92  &\textbf{18.18}  &44.80  &41.32  &\textbf{17.78}  &45.13  &44.20 &\textbf{14.9}  \\
Meta (Ours) &49.17  &39.80  &\textbf{19.30}  &48.36  &40.47  &\textbf{18.63}  &50.98  &43.49  &\textbf{15.61}  \\ \bottomrule \bottomrule
\end{tabular}%
}
\end{table}

\section{\emph{BadSampler} Attack Algorithms}\label{alg:attackalgo}
Here, we provide a detailed version of the pseudocode for the designed \emph{BadSampler} attack algorithm, where Algorithm \ref{algo-2} demonstrates the Top-k sampling attack strategy, and Algorithms \ref{algo-3} demonstrates the meta-sampling attack strategy.
\SetCommentSty{mycommfont}
\begin{algorithm}[htbp]
\DontPrintSemicolon
    \tcc{-- Step 1: Reading Data Index --}
    \Do{complete step 1}{
        read data index from local clients in the form of training samples\;
        get a new data point and add it to a list of unseen data index\;
    }
    \While{local training}{    
    \tcc{-- Step 2: Computing Error Value --}
        Calculate sample difficulty or ${\hat E_D}$ using a surrogate model;\;
        \tcc{-- Step 3: Replacing Sampling Function --}
        \uIf{\textbf{ATK} == "Top-$\kappa$ Sampling"}{
         sort each sample with the sample difficulty\;
         obtain the hard sample pool $\mathcal{H}_{bad}$\hfill $\rhd$\textcolor{gray}{Refer to Eq. \eqref{eq-8}}\;
         perform local training on the adversarial training batch from $\mathcal{H}_{bad}$\;
        }
        \uElse{
        construct meta-state and optimize it using SAC strategy\; 
        obtain the trained adversarial sampler $\xi ( \cdot )$\;
        replace the sampling function and perform adversarial sampling\;
        train the local model on these adversarial training batches \;}   }
    \tcc{-- Step 4: Uploading Model Updates --}
    upload the model updates to the server\;
\caption{Description of the attack steps of the \emph{\textsf{BadSampler}} attack algorithm}
\label{alg-1}
\end{algorithm}
\SetCommentSty{mycommfont}
\begin{algorithm}[htbp]
\DontPrintSemicolon
  \KwInput{Surrogate model $f_s$, the function $get\_index()$ to get the next index of real data, the function $train(f_k, \mathcal{B})$ that trains local model $f_k$ on a batch of data $\mathcal{B}$, and the function $er{r_i}({f_s},{x_i}|D)$ to compute sample's generalization error.}
  \KwOutput{Global model $\omega'$}
    \tcc{Step 1: Reading Data Index}
    data\_index = $[]$\;
    \Do{complete step 1}{
        $I_i = get\_index(x_i)$\;
        
        \uIf{$|\mathcal{I}|$ == $B$}{
            pass batch $\mathcal{B}$ to model $f_s$\;
        }
        \uElse{
            add individual sample index ${I}_{i}$ into data\_index\;
        }
         train($f_s$, $\mathcal{B}$)\;
    }
   \While{local training}{
       \tcc{Step 2: Computing Error Value }
    compute sample difficulty $err_{i}$ via $er{r_i}({f_s},{x_i}|D)$\;
    build a hard sample candidate pool $\mathcal{H}_{bad}$ via Eq. \eqref{eq-8}\;
           \tcc{-- Step 3: Replacing Sampling Function --}
    get adversarial training batch $\mathcal{B}' \sim \mathcal{H}_{bad}$\;
    pass $\mathcal{B}'$ to the local model $f_k$\;
    conduct local training, \ie, $train(f_k, \mathcal{B}')$\;
   }
       \tcc{-- Step 4: Uploading Model Updates --}
    upload the model updates to the server\;
\caption{Top-$\kappa$ sampling attack algorithm} 
\label{algo-2}
\end{algorithm}

\SetCommentSty{mycommfont}
\begin{algorithm}[htbp]
\DontPrintSemicolon
  \KwInput{Surrogate model $f_s$, the function $get\_index()$ to get the next index of real data, and the function $train(f_k, \mathcal{B})$ that trains local model $f_k$ on a batch of data $\mathcal{B}$.}
    \KwOutput{Global model $\omega'$}
    \tcc{Step 1: Reading Data Index}
    data\_index = $[]$\;
    \Do{complete step 1}{
        $I_i = get\_index(x_i)$\;
        
        \uIf{$|\mathcal{I}|$ == $B$}{
            pass batch $\mathcal{B}$ to model $f_s$\;
        }
        \uElse{
            add individual sample index ${I}_{i}$ into data\_index\;
        }
         train($f_s$, $\mathcal{B}$)\;
    }
    \tcc{Step 2: Computing Error Value }
    \While{meta training}{
    train($f_s$, $\mathcal{B}^{\prime}$) at time $t$\;
    use surrogate model $f_s$ to compute ${\hat E_{{D_t}}}$ and ${\hat E_{{D_v}}}$ via Eq. \eqref{eq-9}\;
    build the meta-state via Eq. \eqref{eq-10}\;
    ${\mu _t} \sim \xi ({\mu _t}|{s_t})$\;
    sample $\mathcal{B}'$ via Gaussian function trick \hfill $\rhd$\textcolor{gray}{Refer to Eq. \eqref{eq-11}}\;
    train($f_s$, $\mathcal{B}^{\prime}$) at time $(t+1)$\;
    }
   \While{local training}{
    update the adversarial meta-sampler $\xi ( \cdot )$ via SAC\;
           \tcc{-- Step 3: Replacing Sampling Function --}
    deploy the adversarial meta sampler $\xi ( \cdot )$\;
    get adversarial training batches $\mathcal{B}' \sim \xi ( \cdot )$\;
    pass $\mathcal{B}^{\prime}$ to the local model $f_k$\;
    conduct local training, \ie, $train(f_k, \mathcal{B}^{\prime})$\;
   }
       \tcc{-- Step 4: Uploading Model Updates --}
    upload the model updates to the server\;
\caption{Meta sampling attack algorithm} 
\label{algo-3}
\end{algorithm}

\section{Additional Experiments}\label{appd-3}
\noindent \textbf{IID and Non-IID setting:} \emph{(i)} \textbf{IID:} Here we detail our IID settings. For the CIFAR-10 dataset, we allocate 860 samples to each client on average; that is, a total of 48,000 data samples are used as the training set, and the server stores 12,000 samples as the test set. For the F-MNIST dataset, we allocate 1,020 data samples to each client on average, that is, a total of 56,000 data as the local training set, and the server stores 14,000 samples as the test set. \emph{(ii)} \textbf{Non-IID:} We adopt the classic non-IID setting~\cite{mcmahan2017communication,li2022federated,wang2022fed} where each client can only hold at most two classes of data samples. To do this, we distribute the dataset equally to each client by data class. For example, for client $k$ and CIFAR-10 dataset, its local dataset only has two kinds of data samples with category code ``0'' and ``1''.

\begin{table}[!t]
\caption{Attack impact under the non-IID setting.}
\label{tab-12}
\resizebox{\columnwidth}{!}{%
\begin{tabular}{@{}lccc||ccc||ccc@{}}
\toprule
\toprule
\multicolumn{1}{c}{\multirow{2}{*}{Attacks}} & \multicolumn{3}{c}{$\alpha=0.1$} & \multicolumn{3}{c}{$\alpha=0.5$} & \multicolumn{3}{c}{$\alpha=1$} \\ \cmidrule(l){2-10} 
\multicolumn{1}{c}{} & \begin{tabular}[c]{@{}c@{}}Train \\ acc (\%)\end{tabular} & \begin{tabular}[c]{@{}c@{}}Test\\ acc (\%)\end{tabular} & $\Delta \uparrow$& \begin{tabular}[c]{@{}c@{}}Train \\ acc (\%)\end{tabular} & \begin{tabular}[c]{@{}c@{}}Test\\ acc (\%)\end{tabular} & $\Delta \uparrow$ & \begin{tabular}[c]{@{}c@{}}Train \\ acc (\%)\end{tabular} & \begin{tabular}[c]{@{}c@{}}Test\\ acc (\%)\end{tabular} & $\Delta \uparrow$ \\ \cmidrule(r){1-10}

None &49.14  &47.45  &0  &52.34  &47.45  &0  &54.67  &48.74  &0  \\

Top-$\kappa$ (Ours) &42.85  &28.98  &\textbf{18.47}  &51.88  &32.14  &\textbf{16.60}  &22.82  &22.34 &\textbf{12.81}  \\

Meta (Ours) &37.75  &26.25  &\textbf{21.20}  &48.36  &35.47  &\textbf{11.98}  &52.87  &21.34  &\textbf{27.40}  \\ \bottomrule \bottomrule
\end{tabular}%
\vspace{-0.5cm}
}
\end{table}

\noindent \textbf{Impact of the Non-IID Degree in \emph{\textsf{BadSampler}} Attacks.} Here, we verify the performance of \textsf{BadSampler} on the CIFAR-10 dataset and set the Dirichlet parameter $\alpha=\{0.1, 0.5, 1\}$. The experimental results are shown in Table \ref{tab-12}.

\noindent \textbf{Impact of the Surrogate Models in \emph{\textsf{BadSampler}} Attacks.}
Here, we add a set of additional experiments to explore the impact of different surrogate models on the attack effectiveness of the proposed attack. We set the batch size $B=32$, the number of candidate adversarial training batches $\kappa = 2$, and use FedAvg~\cite{mcmahan2017communication} as AGR. We use the CNN model to conduct classification tasks on the \textsf{CIFAR-10} dataset, and we use the LeNet-5, DenseNet, and AlexNet models as our surrogate models. Table \ref{tab-11} summarizes the experimental results showing that the surrogate model does not significantly affect the performance of the proposed attack.

\section{Why \emph{\textsf{BadSampler}} Work?}\label{appd-4}
Here, we will demonstrate why \emph{\textsf{BadSampler}} is effective from the perspective of training behavior, generalization ability evaluation, and generalization ability visualization.

\noindent \textbf{Training Behavior:} To demonstrate the training behavior of the designed attack, we use the cosine distance of the gradient between clients to represent it. Specifically, we divide the clients into two groups, benign and malicious, and then we calculate the cosine distances of gradients between benign clients and between benign and malicious clients. Next, we give the formal definition of the gradient cosine distance between clients as follows:
\begin{equation}
\mathcal{D}_{\cos }^{i,j} = 1 - \cos \theta  = 1 - \frac{{ < {{\mathbf{g}}_i},{{\mathbf{g}}_j} > }}{{||{{\mathbf{g}}_i}|| \cdot ||{{\mathbf{g}}_j}||}},
\end{equation}
where $\mathbf{g}$ represents the gradient vector of the client. Here, we stipulate that the larger the cosine distance is, the closer the direction of gradient descent between clients is, which means the closer the training behavior is. In the experiment, we conducted five sets of experiments and recorded the mean value of the experimental results of these five sets of experiments in Table \ref{tab-6}. The results show that the training behavior of our attack is not very different from that of benign clients, especially the meta-sampling attack strategy. This means that it is difficult for sophisticated defenders to quickly determine which clients are malicious, which brings new challenges to current advanced defenses.

\begin{figure*}[!t]
    \centering
    \includegraphics[width=1\linewidth]{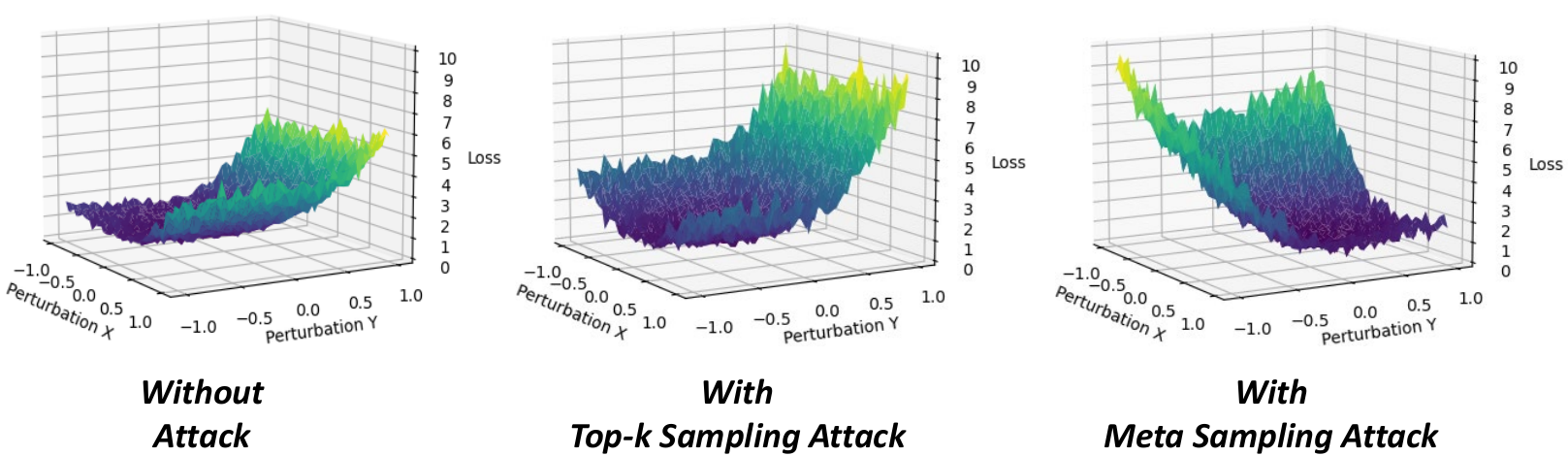}
    \caption{Visual overview of Hessian eigenvalue distributions for benign versus poisoned models.}
    \vspace{-0.4cm}
    \label{fig:my_label}
\end{figure*}

\noindent \textbf{Generalization Ability Visualization:} In the machine learning generalization research domain~\cite{li2018visualizing,mendieta2022local}, top Hessian eigenvalues ($\lambda_{max}$) and Hessian trace ($H_T$) are generally used as key indicators for evaluating model generalization capabilities. In practice, a network with lower $\lambda_{max}$ and $H_T$ is generally a network with stronger generalization ability, that is, the network is less sensitive to small perturbations. This is because lower $\lambda_{max}$ and $H_T$ indicate a more balanced loss space during training and a flatter route to the minimum point. Therefore, we use the parametric loss landscape~\cite{li2018visualizing} visualization with Hessian feature eigenvectors (${\lambda _{ma{x_x}}}$ and ${\lambda _{ma{x_y}}}$) for the resulting global model before and after the attack to show the impact of our attack on the model generalization ability. From Fig. \ref{fig:my_label}, we can clearly see that the loss landscape of the poisoned model changes more drastically and its optimization route is not smooth, even the route optimized to the local optimum is extremely uneven. This means that the generalization ability of the poisoned model is greatly damaged. On the contrary, in the absence of an attack, the loss landscape of the global model changes steadily and its optimization route is smooth and flat.

\noindent \textbf{Generalization Ability Evaluation:} To further quantify the model's generalization error, we follow reference~\cite{mendieta2022local} to compute the difference in Hessian norm ($H_N$) and the difference in Hessian direction ($H_D$) across clients. More specifically, $H_N$ can be used to represent the closeness of model generalization ability between clients and $H_D$ can be used to represent the correlation of training performance. We consider that the model generalization ability of benign clients is generally good, so the above indicators can be used to indirectly quantify the model generalization ability of malicious clients. The formal definition of the above indicators is as follows:
\begin{equation}
    H_N^{i, j}=\left(\left\|\operatorname{Diag}\left(\mathbf{H}_i\right)\right\|_F-\left\|\operatorname{Diag}\left(\mathbf{H}_j\right)\right\|_F\right)^2,
\end{equation}
\begin{equation}
    H_D^{i, j}=\frac{\operatorname{Diag}\left(\mathbf{H}_i\right) \odot \operatorname{Diag}\left(\mathbf{H}_j\right)}{\left\|\operatorname{Diag}\left(\mathbf{H}_i\right)\right\|_F \cdot\left\|\operatorname{Diag}\left(\mathbf{H}_j\right)\right\|_F},
\end{equation}
where $\odot$ is the dot product, ${\operatorname{Diag} \left( {{{\mathbf{H}}_i}} \right)}$ is the diagonal matrix of the Hessian matrix, and $|| \cdot |{|_F}$ is the Frobenius norm. In the experiment, we conducted five sets of experiments and recorded the mean value of the experimental results of these five sets of experiments in Table \ref{tab-6}. From the experimental results, we know that our attack strategy can seriously damage the generalization ability of the model, but the correlation between the training performance of the Top-$\kappa$ sampling strategy and the benign client is relatively different. This may be the reason why the Top-$\kappa$ sampling strategy is easier to defend than the meta-sampling attack strategy.

\section{Proofs}\label{proof}
\subsection{Proof of Lemma \ref{lemm-1}}
\begin{proof}\label{prof-1}
According to Eq. \eqref{eq-5}, we have: $\xi (\omega ) = \sum\limits_{j = 0}^{N - 1} {\sum\limits_{i < j} {\nabla \nabla } } {\ell _j}(\omega ){\ell _i}(\omega )$. If assumption \ref{assum-2} holds, thus, we have:
\begin{align*}
  \mathbb{E}[\xi (\omega )] &= \frac{1}{2}\Bigg(\sum\nolimits_{j = 0}^{N - 1} {\sum\nolimits_{k \ne j} {\nabla \nabla {\ell _j}(\omega )\nabla {\ell _k}} \Bigg)}  \hfill \\
   &= \frac{{{N^2}}}{2}\nabla \nabla F(\omega )\nabla F(\omega ) - \frac{1}{2}\sum\nolimits_{j = 0}^{N - 1} {\nabla \nabla } {\ell _j}(\omega ){\ell _j}(\omega ) \hfill \\
   &= \frac{{{N^2}}}{4}\nabla \Bigg(||\nabla F(\omega )|{|^2} - \frac{1}{{{N^2}}}\sum\nolimits_{j = 0}^{N - 1} {||{\ell _j}(\omega )|{|^2}\Bigg)}.  \hfill 
\end{align*}
\end{proof}

\subsection{Proof of Lemma \ref{lemm-2}}
\begin{proof}\label{prof-2}
Recall that, ${\omega _t} = {\omega _0} - N\eta \nabla \ell ({\omega _0}) + {\eta ^2}\xi ({\omega _0}) + \mathcal{O}({N^3}{\eta ^3})$, thus, we have: $\mathbb{E}({\omega _t}) = {\omega _0} - N\eta \nabla \ell ({\omega _0}) + {\eta ^2}\mathbb{E}(\xi ({\omega _0})) + O({N^3}{\eta ^3}).$
Then we according to Lemma \ref{lemm-2} have:
\begin{align*}
  \mathbb{E}({\omega _t}) &= {\omega _0} - N\eta \nabla F ({\omega _0}) \hfill \\
   &+ \frac{{{N^2}{\eta ^2}}}{4}(||\nabla F ({\omega _0})|{|^2} - \frac{1}{{{N^2}}}\sum\nolimits_{j = 0}^{N - 1} {||} {\ell _j}({\omega _0})|{|^2}) + O({N^3}{\eta ^3}) \hfill .
\end{align*}
\end{proof}

\subsection{Proof of Theorem \ref{theo-1}}
\begin{proof}\label{prof-1}
First, we expand the error term expectation as follows: $\mathbb{E}[||\nabla F({\omega ^*}) - \nabla F({\omega ^b})|{|^2}] = \frac{1}{{{B^2}}}\mathbb{E}\bigg(\sum\limits_{i = 1}^B {\sum\limits_{j = 1}^B {(\nabla {F_i}({\omega ^*}) - \nabla F({\omega ^*})) \cdot } } (\nabla {F_j}({\omega ^p}) - \nabla F({\omega ^p}))\bigg)$. To keep the notation clean, we let ${X_i} = (\nabla {F_i}({\omega ^*}) - \nabla F({\omega ^*}))$ and ${X_j} = (\nabla {F_i}({\omega ^p}) - \nabla F({\omega ^p}))$. Thus, we have:
\begin{align*}
  \mathbb{E}[||\nabla F({\omega ^*}) - \nabla F({\omega ^b})|{|^2}] &= \frac{1}{{{B^2}}}\mathbb{E}(\sum\limits_{i = 1}^B {\sum\limits_{j = 1}^B {({X_i} \cdot {X_j})} } ) \hfill \\
   &= \frac{1}{B}\mathbb{E}({X_i} \cdot {X_j}) + \frac{{B - 1}}{B}\mathbb{E}({X_i} \cdot {X_{j \ne i}}) \hfill
\end{align*}
Since Lemma \ref{lemm-1} and Lemma \ref{lemm-2} hold, we have:
\begin{align*}
    \mathbb{E}[||\nabla F({\omega ^*}) - \nabla F({\omega ^b})|{|^2}] = \frac{1}{{NB}}\sum\limits_{i = 1}^N {{X_i} \cdot {X_i}}  + \frac{{(B - 1)}}{{BN(N - 1)}}\sum\limits_{i = 1}^N {\sum\limits_{j \ne i} {{X_i} \cdot {X_j}} } 
\end{align*}
Next, we recall that $\sum\nolimits_{i = 1}^N {{X_i} = \sum\nolimits_{i = 1}^N {\bigg(\nabla {F_i}(} } {\omega ^*}) - \nabla F(\omega )\bigg) = 0$, thus we have:
\begin{align*}
  \mathbb{E}[||\nabla F({\omega ^*}) - \nabla F({\omega ^b})|{|^2}] &\leqslant \frac{1}{{NB}}\sum\limits_{i = 1}^N {{X_i} \cdot {X_i}}  - \frac{{(B - 1)}}{{BN(N - 1)}}\sum\limits_{i = 1}^N {{X_j} \cdot {X_j}}  \hfill \\
  & = \frac{1}{{NB}}(\sum\limits_{i = 1}^N {{X_i} \cdot {X_i}}  - \frac{{B - 1}}{{N - 1}}\sum\limits_{i = 1}^N {{X_j} \cdot {X_j}} ). \hfill 
\end{align*}
If Assumptions \ref{assum-1}--\ref{assum-2} hold, then $\sum\nolimits_{i = 1}^N {{X_i} \cdot {X_i}}  \leqslant N{(1 - \frac{1}{N})^2}\varepsilon _0^2$ and $\sum\nolimits_{i = 1}^N {{X_j} \cdot {X_j}}  \leqslant N{(1 - \frac{1}{N})^2}\varepsilon _b^2$, thus we have:
\begin{align*}
\mathbb{E}[||\nabla F({\omega ^*}) - \nabla F({\omega ^b})|{|^2}] \leqslant B{(1 - \frac{1}{N})^2}\varepsilon _0^2 - \frac{{B - 1}}{{NB}}\varepsilon _b^2.
\end{align*}
\end{proof}

\end{document}